# EXAMINATION OF WAVE-PARTICLE DUALITY VIA TWO-SLIT INTERFERENCE

Mario Rabinowitz
Armor Research, lrainbow@stanford.edu
715 Lakemead Way, Redwood City, CA 94062-3922

**Abstract**

The wave-particle duality is the main point of demarcation between quantum and classical physics, and is the quintessential mystery of quantum mechanics. Young's two-slit interference experiment is the arch prototype of actual and gedanken experiments used as a testing ground of this duality. Quantum mechanics predicts that any detector capable of determining the path taken by a particle through one or the other of a two-slit plate will destroy the interference pattern. We will examine both the experimental and theoretical attempts to test this assertion, including a new kind of experiment, and to grasp the underlying truth behind this mystery from the earliest days to the present. Where positions differ, the views of both sides are presented in a balanced approach.

PACS Nos.: 03.65.Bz, 42.25.Hz, 3.65.-w, 01.65.+g, 01.70.+w.

## 1. Introduction

Little could Newton and Huygens foresee that their eighteenth century debate about whether light is particle-like or wave-like[1] would foreshadow a debate about the wave-particle duality of both light and matter that is still being waged. Little did Newton, Huygens, Young, and their contemporaries know that the 1802 victory of Thomas Young's two-slit interference experiment[2] in favor of the wave nature of light was only a virtual victory, and that his two-slit experiment in a myriad of variations would be the paradigm to this day for illustration and consideration of the issues related to the wave-particle duality.

Though in retrospect he may not have liked it, Einstein was reponsible for initiating the wave-particle duality in 1905 by associating the concept of an indivisible particle, the photon, with the well-established wave nature of light.[3] He assigned the discrete energy $h\nu$, to a photon ($h$ is Planck's constant, and $\nu$ is the frequency of the light) to explain the photoelectric effect for which he was awarded the Nobel prize in 1921. Einstein described light[3] as "consisting of a finite number of energy quanta which are localized at points in space, which move without dividing, and which can only be produced and absorbed as complete units." This view has remained unchanged since its inception. In 1923, de Broglie inverted the concept to associate a wave with particles.[4] The paradigm shift for all this started in 1900, when Planck explained the blackbody radiation curve in terms of harmonic oscillators which could absorb or emit energy only in discrete units $h\nu$.[5]

Ever since the beginnings of quantum mechanics, the wave-particle duality has captured the imagination of physicists. It took only four years after Einstein explained the photoelectric effect by quantization of light for the first single-photon interference experiment to be performed. Taylor did it in 1909 with a flame light source, a diffraction grating and a photographic plate.[6] Taylor achieved the penultimate (considering delayed choice or quantum erasure to be the ultimate) double-slit experiment, in which particles are directed one at a time at a grating. These single particles are each diffracted in passing through the slit(s). Each particle produces one displaced spot on a screen opposite the slits; and with the collection of a large number of particles an interference pattern emerges. The term luminescent screen shall be used herein for anything capable of single particle detection such as an array of photomultipliers.

## 2. Two-Slit Interference Gedanken Experiments

### 2.1 *Einstein-Bohr debates*

In one of the first of the epoch-making debates between Albert Einstein and Neils Bohr, the two-slit experiment was a pivotal point of contention at the 1927 fifth Solvay Congress in Brussels.[7] Three versions of the two-slit recoil gedanken (thought) experiment are attributed to Einstein. In one of them, he pointed out that when a photon is diffracted to the right of the interference maximum, it imparts a greater momentum transfer to the slit plate if it goes through the left slit than if it goes through the right slit because it acquires a larger transverse momentum in going through the left slit. Thus, in principle, by observing the recoil of the slit plate, one may determine the traversed slit. However, Bohr[8] showed that for such an observation to leave the interference pattern undisturbed, the uncertainty principle would have to be violated for the slit plate. Otherwise the recoiled slit plate would move so much each time that the interference pattern would be washed out. Let us look in detail at Bohr's arguments.

**2.1.1** *Two-Slit Plate Recoil*

Bohr's contention goes roughly as follows. Consider the second maximum to the right of and spaced a distance Δx from the central maximum. Particles detected on the screen at this position have different momenta $p_\ell$ and $p_r$ depending on which slit they came from because of the greater angle of deflection from the left slit than the right slit. The difference of the x components of momentum is

$$\Delta p_x \approx \frac{h}{\Delta x}. \tag{1}$$

(If this is not clear, it will become clear after the more rigorous and lengthy derivation which follows immediately. Equation (1) corresponds to Eq. (16) with $\Delta x = \lambda D/s$.) Therefore the momentum transfer to the slit plate must be measured more accurately than the expression of Eq. (1) to determine the slit of passage. The requires the measured error

$$\delta p_x < \Delta p_x \approx \frac{h}{\Delta x}. \tag{2}$$

The uncertainty of measuring the impact spot must be less than half the fringe spacing

$$\delta x < (\text{fringe spacing})/2 = \Delta x/2. \tag{3}$$

Multiplying Eqs. (2) and (3) together, yields

$$\delta p_x \delta x < \frac{h}{\Delta x}\left(\frac{\Delta x}{2}\right) = h/2. \tag{4}$$

This appears to beat the uncertainty principle

$$\delta p_x \delta x \geq h/4\pi = \hbar/2 \tag{5}$$

by a factor of 2π. So let us look at this a little more carefully.

The magnitude of the (transverse) x component of momentum of a particle coming from the right slit is

$$p_{xr} = p\left(\begin{matrix}\text{sine of}\\ \text{deflection angle}\end{matrix}\right) = \frac{h}{\lambda}\left(\frac{x - s/2}{\left[D^2 + (x - s/2)^2\right]^{1/2}}\right) \approx \frac{h}{\lambda}\left(\frac{x - s/2}{D}\right), \tag{6}$$

where $\lambda = h/p$ is the de Broglie wave length. Somewhat as shown in the lower half of Fig.1, s is the separation of the slits, D is the distance from the slit plate to the interference screen, and x is the position of the spot on the screen measured from the center line. Similarly for a particle from the left slit

$$p_{x\ell} = \frac{h}{\lambda}\left(\frac{x + s/2}{\left[D^2 + (x + s/2)^2\right]^{1/2}}\right) \approx \frac{h}{\lambda}\left(\frac{x + s/2}{D}\right). \tag{7}$$

In order to determine which slit the particle traversed, we must be able to measure the momentum of the slit plate with an uncertainty

$$\Delta p_x < p_{xr} - p_{x\ell} \approx \frac{h}{\lambda}\left(\frac{s}{D}\right). \tag{8}$$

Let us tighten up the requirement on our ability to discriminate interference fringes compared to the criteria used for Eq. (3):

$$\Delta x < (\text{fringe spacing})/2\pi = \lambda D/(2\pi s). \tag{9}$$

From Eqs. (8) and (9), we require the product of the measurement uncertainties to be

$$\Delta p_x \, \Delta x < \frac{h}{\lambda}\left(\frac{s}{D}\right)\left(\frac{\lambda D}{2\pi s}\right) = \hbar. \tag{10}$$

Again, this appears to barely comply with the uncertainty principle by a factor of 2, though it is a little too close for comfort (cf. last sentence of Section 2.1.3).

**2.1.2** *Single-Slit Source Plate Recoil*

Another version is recalled by Bohr in which Einstein suggests detection of the momentum transfer to a single-slit plate between the light source and the two-slit plate, with a similar violation of the uncertainty principle.[7] Bohr[9] argued that if $\omega$ is the angle between the single slit and the two slits, for the conjectured paths of a particle through the left or right slits, the difference of momentum transfer in these two cases is

$$\Delta p \approx p\omega \approx \left(\frac{h}{\lambda}\right)\left(\frac{s}{D}\right), \tag{11}$$

where D is the distance from the single-slit plate to the two-slit plate, and also from the two-slit plate to the interference screen. Bohr points out that the same result will also ensue if the two-slit plate is not midway between the single-slit plate and the screen. The uncertainty principle requires an uncertainty in position of the single-slit plate

$$\Delta x > \left(\frac{\hbar}{2\Delta p}\right) = \left(\frac{\hbar\lambda D}{2hs}\right) = \left(\frac{\lambda\!\!\!^{-} D}{2s}\right) \tag{12}$$

causing the same uncertainty in the position of the fringes. Since this is also approximately the fringe separation $\left(\frac{\lambda D}{s}\right)$, Bohr argued that no interference effect can appear. Since $\Delta x$ is a factor of $4\pi$ smaller than the fringe separation, it seems possible that something resembling an interference pattern could survive, were it not for the proviso of the last sentence of Section 2.1.3.

Bohr[9] recalls, "but, in spite of all divergencies of approach and opinion, a most humorous spirit animated the discussions. On his side, Einstein mockingly asked us whether we could really believe that...'ob der liebe Gott wurfelt' [does God play dice?].... I remember also how at the peak of the discussion Ehrenfest, in his affectionate manner of teasing his friends, jokingly hinted at the apparent similarity between Einstein's attitude and that of the opponents of relativity theory; but instantly Ehrenfest added that he would not be able to find relief in his own mind before concord with Einstein was reached."

**2.1.3** *Interference Screen Recoil*

An equally ill-fated third version has Einstein proposing that the lateral kick imparted by a photon to the interference screen could be used to determine which slit the photon traversed as it went to the screen. To record the interference fringes, the luminescent screen location must be fixed within a lateral displacement

$$\Delta x < (\text{fringe spacing}) / 2\pi = \lambda D / (2\pi s), \tag{13}$$

where s is the slit spacing, D is the distance of the source from the slit-plate. The x axis lies on the geodesic joining the two slits on the slit plate, and the y axis is on the center line from the source to the slit plate. Although Fig. 1 is intended to illustrate a different, new kind of experiment, the lower portion can be helpful in indicating the geometry here, which is common to all the two-slit interference experiments.

If the particle comes to the center of the central maximum from the left slit, its momentum is

$$\vec{p}_\ell = p_{x\ell}\mathbf{i} + p_{y\ell}\mathbf{j} + p_{z\ell}\mathbf{k} = p\left(\frac{s}{2D}\right)\mathbf{i} + p_{y\ell}\mathbf{j} + p_{z\ell}\mathbf{k}, \tag{14}$$

where **i**, **j**, and **k** are the unit vectors in the x, y, and z directions. If it comes to the central maximum from the right slit

$$\vec{p}_r = -p_{xr}\mathbf{i} + p_{yr}\mathbf{j} + p_{zr}\mathbf{k} = -p\left(\frac{s}{2D}\right)\mathbf{i} + p_{yr}\mathbf{j} + p_{zr}\mathbf{k}. \tag{15}$$

To ascertain the slit through which the particle came, one must determine the transverse momentum imparted to the screen by distinguishing between the particle's momentum if it comes from the left or right slit, i.e. the x components of momentum must be distinguishable. It is not necessary to assume that $p_{y\ell} = p_{yr}$, and $p_{z\ell} = p_{zr}$, but only that $p_\ell = p_r$. Eqs. (14) and (15) imply that in order to discriminate, the measurement of the x component of the screen momentum, it must be made with an uncertainty less than the difference between these two possible x components of momentum of the particle

$$\Delta p_x < p\left(\frac{s}{D}\right) = \frac{h}{\lambda}\left(\frac{s}{D}\right). \tag{16}$$

The quantity on the right hardly changes if we consider a side maximum of the interference pattern rather than the center of the central maximum as can be seen from Eqs. (6), (7) and (8). From Eqs. (13) and (16)

$$\Delta p_x \Delta x < \frac{h}{\lambda}\left(\frac{s}{D}\right)\left(\frac{\lambda D}{2\pi s}\right) = \left(\frac{h}{2\pi}\right) = \hbar, \tag{17}$$

which misses violating the uncertainty principle by a factor of 2.

Since the details of the interaction of the particle with the screen are not included, i.e. whether the particle is absorbed or partially reflected at the point of impact, the inherent uncertainty of the screen position due to thermal motion, etc., one may consider that the uncertainty principle is violated for the screen in all the above cases in trying to obtain simultaneously both particle and wave information.

**2.2 *Feynman's impossibility illustration***

In discussing the wave-particle duality in connection with the two-slit interference experiment, Richard Feynman[10] said, "We choose to examine a phenomenon which is impossible, *absolutely* impossible, to explain in any classical way, and which has in it the heart of quantum mechanics. In reality, it contains the only mystery." To illustrate this point, he presented a gedanken experiment in which a light is placed behind the slit plate between the two slits. Feynman argues that an electron shot at the slit plate will scatter light in the vicinity of the hole through which it passed, i. e. nearer that hole than the other.[10] He points out that for a sufficiently small wavelength of light to make this distinction, the electron would be scattered too much to produce an interference pattern. For a sufficiently soft photon that would not disturb the interference pattern, its wavelength would be too long to discriminate from which slit the electron came.

**2.3 *Wooters and Zurek inaccurate determinaion***

Wooters and Zurek[11] propose to modify the double-slit experiment in such a way that "one can retain a surprisingly strong interference pattern by not insisting on a 100% reliable determination of the slit through which each photon passes." They separately make measurements of momentum and position of the single-slit plate (as in the Einstein-Bohr debate of Section 2.1.2) and ask the question: "Does our choice of what to measure affect the total interference pattern? Do we not violate the complementarity principle by measuring both the fringes and the kick?" In addition to analyzing their version of this Einstein gedanken experiment, they propose a multiplate double-slit interference experiment which they think can be done in practice to illustrate their theoretical findings and the possibility of "delayed choice."

They divide up the ensemble of measurements into subensembles depending on which kind of measurement they made. They find that although the partial interference patterns depend on the kind of measurement performed, the total interference pattern related to the total ensemble of measurements and hence the sum of the interference patterns is always the same. Their momentum measurements of the single-slit plate result in "*smeared out* but *centered* " partial interference patterns. The position measurement of the single-slit plate yields "*perfect* but *shifted* partial interference patterns." They conclude, "The more clearly we wish to observe the wave nature of light, the more information we must give up about its particle properties."

They point out a similarity to the Einstein-Rosen-Podolsky[12] paradox in that their measurement on the single-slit plate can be made after the photon has interacted with it. They ponder this delayed choice option they present, in asking: "How does the photon know in which partial interference pattern to fall? How can it know what we decided to measure when it is already separated from the plate by a large distance?" We all would like to know the answers to these questions. As presented these experiments are gedanken (thought) experiments that are consistent within the framework of quantum mechanics. Further deliberation is needed to determine this consistency within the framework of nature.

**2.4** ***Dominance of complementarity over the uncertainty principle?***

Scully et al[13] argue that although prohibition of the simultaneous observation of wave and particle behavior is often enforced by the uncertainty principle's momentum-position relation, this complementarity of wave and particle properties is more basic and may also be enforced by other mechanisms. They point out that although complementarity is usually associated with the wave particle duality of matter, it is a much more general quantum mechanical concept. For them in a dynamical system, for each degree of freedom there are pairs of complementary observables where precise measurement of one makes the outcome of a measurement of the other completely unpredictable. Thus they put complementarity at a more fundamental level than the uncertainty principle, including the canonically conjugate variables of the uncertainty principle as a subset. They say that "the actual mechanisms that enforce complementarity vary from one experimental situation to another." In their examples they allow one enforcer to be the momentum-position relation of the uncertainty principle. This appears to be an inconsistency in their point of view, since one may expect the enforcers of complementarity to be more fundamental than complementarity.

Nevertheless, in the realm of the wave-particle duality Scully et al contend that complementarity may be at work even when the uncertainty principle is not. They base this contention on intriguing gedanken experiments with atom interferometers using new modern quantum optic detectors. They say, "we find that the interference fringes disappear once we have which-path information, but we conclude that this disappearance originates in correlations between the measuring apparatus and the systems being observed. The principle of complementarity is manifest although the position-momentum relation plays no role."

In their astutely proposed experiments, their detectors are two micromaser cavities in front of each slit of the traditional two-slit interference experiment. They can determine the path of an atom through one micromaser cavity or the other in front of the two slits, because an atom in passing through a cavity will make a transition from an excited state to a lower state because of the interaction with the photons in the cavity. For them the wave function consists of two components in which the first is affected by a weak attractive potential, and the second to an equally weak repulsive potential. These potentials affect the internal atomic transition accompanied by the emission of a photon. There is no net momentum transfer to the atom during its interaction with the cavity fields, as the atom regains its same initial momentum after traversing the cavity. This suggested experiment would be so delicate that one would not expect the interference pattern to be disturbed. Yet surprisingly, they expect the interference pattern to be destroyed if they make an observation on which cavity the atom passed through. More amazingly, in a suggested variation of this proposed experiment they claim that the interference effects could either be destroyed or restored by manipulating the cavities long after the atoms have passed. Such "quantum erasure" claims are closely related to the "delayed choice" claims of others.

Two clarifications of their quantum erasure and quantum optical tests of complementarity need to be made. Both in this paper,[13] and in a recent popularized version in Scientific American,[14] unless one reads these articles very carefully one is left with the impression that these experiments have already been done. Reading their prior papers,[15-19] one finds that these experiments have yet to be done. Even with careful reading of these papers,[13, 14] not the least hint is given of a number of scientific papers with opposing views, and one is left with the impression that their views on complementarity and quantum erasure are completely accepted by the scientific community. As we shall see in the next and other following sections, they have been challenged by their colleagues on the simpler claim that complementarity is an independent pillar of quantum mechanics, rather than merely a consequence of the uncertainty principle. Quantum erasure may also be challenged.

**2.5 *Dominance of the uncertainty principle over complementarity?***

**2.5.1** *Storey, Tan, Collett, and Walls*

Although quantum mechanics declares that any detector capable of determining the path taken by a particle in a double slit experiment will destroy the interference pattern because of the complementarity principle, the physical mechanism is not given by which the interference is destroyed. Storey et al[20] disagree with Scully et al[13-19] who suggested that complementarity must be accepted as an independent component of quantum mechanics. Storey et al argue that complementarity is simply a consequence of the uncertainty principle, and that the Scully et al scheme is in principle no different than the Einstein recoiling slit.

Storey et al's paper is directed at showing that for these suggested experiments "the loss of interference from a double slit in the presence of a *welcher Weg* [which path] detector is physically caused by momentum kicks, the magnitude of which are determined by the uncertainty principle." Thus they conclude that in their proffered experiment, Scully and his colleagues overlooked the successive momentum transfers due to the recurring emission and absorption of photons by an atom in going through a given cavity. For Storey et al, this is what wipes out the interference fringes. Earlier Tan and Walls[21] came to the same conclusion in 1993 in arguing against Scully et al.

**2.5.2** *Bhandari*

By reversing the usual approach, Bhandari[22] deduces the presence of a geometric phase from the assumption that the interference fringes must disappear as a result of the "which path" determination in a two-slit experiment. (Berry[23] discovered the geometric phase in the context of the adiabatic evolution of quantum systems. This was generalized to non-adiabatic evolution by Aharanov and Anandan.[24]) Bhandari starts with the hypothesis that the destruction of the interference pattern is due to a randomization of the phase of the interfering waves as a result of the "which path" determination. In looking for the source of the random phase, he finds it to be a geometric phase. An inversion of Bhandari's argument would seem to imply that if a geometric phase is not present then the "which path" determination was not made.

He considers a variation of the Einstein slit-plate recoil gedanken experiment. A right-hand circularly polarized beam of light is split by a beam splitter into two beams, each of which passes through a half-wave plate (HWP), which are then recombined after travelling separate paths. In passing through the HWP, each photon reverses its helicity (its spin angular momentum), and in the process imparts a net angular momentum $2\hbar$ to the HWP. Determination of this angular momentum change of the HWP by detectors in front of each of the two slits would provide the "which path" information.

Bhandari concludes that the interference pattern is lost for a similar reason to that in the Einstein-Bohr[7] debate, and that the geometric phase is responsible for losing angular momentum information (which could otherwise determine the path) when the interference pattern is retained. Bhandari further concludes that there is no difference in principle between the gedanken experiment of Einstein and the proposed experiments of Scully et al[13-19] to detect a photon emitted by an atom passing through a maser cavity. He says, "(1) Einstein's experiment and the one proposed in this paper are exactly similar except for the replacement of the conjugate variables x [position] and p [momentum] in Einstein's proposal with the variables $\varphi$ [angle] and $L_z$ [component of the angular momentum of the HWP in the direction of the beam axis] in the present one. (2) The [gedanken] experiment of Scully and Walther[17] is exactly similar to our proposed experiment except for the replacement of the conjugate variables $L_z$ and $\varphi$ in the latter with the pair N, $\Theta$ in the former, where N stands for the number of energy quanta and $\Theta$ for the phase of the oscillating cavity mode." He shows that the uncertainty relation in his proposed experiment, $\Delta L_z \Delta\phi \geq \hbar / 2$, plays the same role as the uncertainty relation $\Delta N \Delta\Theta \geq 1/2$ in the proposed experiments of Sculley et al[13-19].

**2.6 *Wheeler's delayed choice***

The distinguished physicist John Archibald Wheeler[25] restructured the third version in Section 2.1 of the Einstein detection scheme so that optical discrimination of the interference pattern is detected rather than photon momentum transfer. In Wheeler's proposed experiment, the experimenter may exercise his discretion of whether to determine which slit the photon chose, or to build up the interference pattern -- after the photon went through the slit. In Wheeler's words: "But the essential new point is the timing of the *choices* -- between observing a two-slit effect and a one-slit one -- until after the single quantum of energy in question has *already* passed through the screen....Then let the general lesson of this apparent time inversion be drawn: 'No phenomenon is a phenomenon until it is an observed phenomenon.' In other words, it is not a paradox that we choose what *shall* have happened after 'it has already happened.' It has not really happened, it is not a phenomenon, until it is an observed phenomenon."

Wheeler describes a two-slit interference experiment in which the choice of measuring the direction of the particle or its wavelike nature at the screen is made after the particle has been detected. He places an interference screen at the focal plane of a lens at a distance L from the slits, where a Fraunhofer pattern can be observed. If instead, one wants to determine a given photon's trajectory, the screen is turned aside and the photon activates one photomultipier or another. Wheeler goes on to say, "To be forced to choose between complementary modes of observation is familiar, but it is unfamiliar to make this choice after the relevant interaction has already come to an end. Moreover, one can assert this 'voice in what should have happened, after it appears to already have happened' in illustrations of complementarity other than the double slit, by suitable modification of the idealized apparatus."

Wheeler also presents delayed choice variations of other classic gedanken quantum experiments. His version of the gamma ray microscope of Heisenberg[26] and Bohr[8] is illuminating. A lens of angular opening $\varepsilon$, brings to a focus light of wavelength $\lambda$within an uncertainty in position $\Delta x \sim \lambda / 2\pi\varepsilon = \bar{\lambda} / \varepsilon$. The photon scattered into the lens gives the electron a lateral kick, with momentum uncertainty

$$\Delta p \sim \begin{bmatrix} \text{photon} \\ \text{momentum} \end{bmatrix} \begin{bmatrix} \text{angular opening} \\ \text{of the lens} \end{bmatrix} \sim \left( \frac{\hbar}{\bar{\lambda}} \right) \varepsilon .$$

This is in accord with the uncertainty principle. However, Wheeler points out that here also a delayed choice would determine whether we observe the particle or wave nature of light. "However, the uncertainty in the lateral kick can be reduced to a very small fraction of ... [the above value] by placing a sufficiently great collection of sufficiently small photodetectors at a little distance above the lens. Whichever one of them goes off signals the direction of the scattered photon and thus the momentum imparted to the electron." Wheeler then goes on to add the feature and puzzling consequences of delayed choice by deciding which set of photodetectors to activate "after the lens has *already* finished transmitting the photon." Needless to say, such results would be bewildering to most people.

**2.7 *Necromancy of quantum mechanics?***

In contrast to Wheeler, the distinguished physicist Edwin Jaynes[27] has presciently pointed out "that present quantum theory not only does not use -- it does not even dare to mention -- the notion of a 'real physical situation.' Defenders of the theory say that this notion is philosophically naive, a throwback to outmoded ways of thinking, and that recognition of this constitutes deep new wisdom about the nature of human knowledge. I say that it constitutes a violent irrationality, that somewhere in this theory the distinction between reality and our knowledge of reality has become lost, and the result has more the character of medieval necromancy than of science. It has been my hope that quantum optics, with its vast new technological capability, might be able to provide the experimental clue that will show us how to resolve these contradictions." The strict definition of necromancy is devination by pretended communication with the spirits of the dead. The most generous view of Jaynes' use of this word is that the philosophy of quantum mechanics is akin to sorcery.

Jaynes uses a number of examples to support his view. He points out that one of the most enduring incorrect beliefs about quantum electrodynamics is that it endows light by means of the photon to wipe out the interference effects of classical electromagnetic theory. In 1954, an experiment was proposed[28] to observe interference between Zeeman components of a spectral line. This possibility was denied by quantum theorists because as Jaynes points out satirically, "as everybody knows, 'a given photon interferes only with itself.' Yet the photoelectric klystron worked, the beats [interference] were seen, and an important lesson was learned about the meaning and correct application of quantum theory." However, the lesson was soon lost. An experiment was proposed[29] to measure stellar diameters by interference measurements involving fourth order spatial correlation functions of the field. Again the possibility of the effect was denied by eminent theorists. Yet the experiment worked as predicted by classical electromagnetic theory. But the lesson was lost again. In the early 1960's, shortly after the invention of the laser, theorists said "that it is fundamentally impossible to observe beats between independently running lasers --a given photon interferes only with itself. Jaynes notes that the beats appeared on schedule -- just as classical electromagnetic theory predicted.

It is not that Jaynes is opposed to quantum theory in general, or quantum electrodynamics in particular. His opposition seems more to be concerned with the improper interpretation and application of quantum theory.

**2.7** ***Non-locality of a single photon***

Following the earlier lead of Tan, Walls and Collett,[30] and Oliver and Stroud,[31] Hardy[32] proposes an experiment to demonstrate the non-locality (wave-like nature) of a **single** photon in which it appears to be in two places at the same time. He even presents such a persuasive argument that this leads to a contradiction that "one might be tempted to think that quantum mechanics must be wrong." However, he goes on to show that there is an implicit assumption of locality in this argument, and without this assumption there is no contradiction. This proposed non-locality experiment illustrates the contrafactual nature of quantum mechanics, in that the possibility of following a path has essentially the same effect as if the path were actually taken.

Hardy points out that "as early as 1927 ... [at the Fifth Solvay Conference[7]], Einstein presented the collapse of a single particle wave packet to a near position eigenstate as a paradigm for nonlocality in quantum mechanics (indeed one might even say that he anticipated the Einstein-Podolsky-Rosen[12] argument in the context of this example)." It is clear that the original concerns of Einstein in 1927 about the nonlocality of a single particle in quantum mechanics are indeed justified by both theoretical and experimental findings.[33-35]

Hardy addresses the question of whether this nonlocality is also true of other particles or just restricted to photons as in his derivation. His answer is that an analogous proof "could be constructed for any type of particle for which it is possible to prepare a direct superposition of that particle with the vacuum. However, for a vast range of types of particles there are superselection rules that prohibit just exactly this, and nonlocality with single particles of this type could not be observed." This may support the view that the notion of particle for a photon may be quite different than for most other particles, though photons sometimes exhibit particle-like properties. Yet as in so many other instances, it would not be surprising if the wave-particle duality in all its manifestations holds across the board for light and all particles alike.

## 3. Two-Slit Experiments Without the Slits

### 3.1 *Interference of light scattered from two atoms*

Eichman et al[36] scatter light of 0.194μ wavelength from two $^{198}Hg^{+}$ ions trapped by Doppler laser cooling at separations of 3.7, 4.3, and 5.4μ. The Hg ions are the analog of the slits in the Young's two-slit interference experiment. Light, both elastically and inelastically, scattered from the two ions produces an interference pattern. Quantum mechanics predicts that interference must result in the scattered light when it is not possible at each collision to determine which ion scattered the photon. When the spectator atom is distinguished from the scattering atom, thus determining the photon path, the interference pattern is lost.

They contend that by exploiting the atom's internal level structure they showed that determination of the scattered photon's trajectory obliterates the interference pattern, without violation of the position-momentum uncertainty relation. For them, it would be incumbent for those who would not agree, to prove that the enforcer of complementarity in this case is the uncertainty principle. In any case the result of their extraordinary experiment appears incontrovertible.

They obtain a particle-like behavior when they determine the photon trajectory, which begins at the source, intersects one of the atoms, and continues to the detector without producing an interference pattern. When they don't determine the photon path, they obtain a wavelike behavior producing a typical two-slit diffraction pattern. They use polarization-sensitive detection of the scattered photons to switch on either the wave-like or the particle-like character of the scattered photon.

**3.2** ***Interference for two atoms radiating a single photon***

P. Grangier et al[37] present experimental evidence for a modulation in the time-resolved atomic fluorescence light following the photodissociation of $Ca_2$ molecules. They conclude that "This modulation is due to an interference effect involving two atoms recoiling in opposite directions, while only one photon is emitted. ... This experiment is thus analogous to a single-photon Young's [double] slit experiment, in which the 'slits' (the atoms) are moving." They point out that one might hope to determine the photon trajectory by observing the momentum of each atom after the photon is detected. This would lead to knowing which atom (slit) was the emitter since it received the extra momentum hν/c from the fluorescence photon.

The relevant quantity is the difference of momenta of the two atoms, measured with an uncertainty less than hν/c. However if the momentum uncertainty is less than hν/c, the uncertainty principle implies that the uncertainty in the relative position of the two atoms must be greater than λ, causing the interference effect to be lost. They conclude that in their experiment, the initial dispersion of the relative position of the atoms is very small compared to λ, while the momenta difference dispersion is greater than hν/c for the atoms with recoil velocity of ≈ 500 m/sec, so that there is no way to know "which atom emitted the photon".

**4. A New Kind of Experiment**

All the founders of quantum theory were at the fifth Solvay Congress, from Planck, Einstein, and Bohr to de Broglie, Heisenberg, Schrodinger, and Dirac, to witness debates which continued into the next Congress. As Rosenfeld[38] relates, "At the sixth Solvay Conference, in 1930, Einstein thought he had found a counterexample to the uncertainty principle. It was quite a shock for Bohr ... he did not see the solution at once. During the whole evening he was extremely unhappy, going from one to the other and trying to persuade them that it couldn't be true, that it would be the end of physics if Einstein were right; but he couldn't produce any refutation. I shall never forget the vision of the two antagonists leaving the club [of the Fondation Universitaire]: Einstein a tall majestic figure, walking quietly with a somewhat ironical smile, and Bohr trotting near him, very excited.... The next morning came Bohr's triumph."

The case against determination of both concurrent trajectory and wave properties of a particle has appeared so strong that ever since the Einstein-Bohr debate at the fifth Solvay Congress of 1927 on the Einstein recoil experiment,[7-9] no new gedanken or experimental challenges have arisen. Instead the theoretical and experimental efforts have been directed at further confirmation of the established view. The present situation is not at all reminiscent of the lively Einstein-Bohr debates. In the hope of both gaining new insight into a vitally important issue, and of reviving the spirit of the old debates, we have proposed a novel gedanken experiment that does not violate the uncertainty principle, in which manifestly the traversed slit can be apprpximately determined by inference without disturbing the slit plate, the particle, the interference screen, or destroying the interference pattern.[39] We are well-aware that previous attempts to unravel the wave-particle duality have always been shown to have some fatal flaw. Yet, it was widely thought that tunneling could only be understood quantum mechanically until we showed that not only is there classical tunneling, but that it can be closely related to quantum tunneling.[40]

Let us start with an emission source of particle pairs, much the same as in the Einstein-Podolsky-Rosen[12] gedanken experiment and as modified by Bohm[41]. The source is on the center line axis of symmetry between a screen A and a double-slit plate, followed by a screen B, as shown in Figure 1. The source here is either point-like, i.e. small compared with the other dimensions of the apparatus, or an extended source that emits symmetrically with respect to its center. An extended source is shown (from which the particle-pair trajectories are in line with the source center) to illustrate that this yields the same results as a point-like source. The distance from the source to the slit plate $D_B$ is >> $\lambda$, the wavelength associated with the particle, so that an approximately plane wave is achieved.

Assume that a quasi-stationary source emits a pair of particles in opposite directions by conservation of momentum, in which ideally the source had or gains little or no momentum perpendicular to the particles' trajectory. When this is not realized in practice, a partial correction can be computed taking into account the observed transverse component of momentum of the source. The source may emit a pair of photons resulting from a radiative cascade of Ca as in the experimental realization of the Einstein-Podolsky-Rosen-Bohm (EPRB) gedanken experiment by Aspect et al , and in its modifications.[33,34,35] The 2-$\gamma$ decays produced by ground-state quasi-stationary positronium annihilation could also be used in principle. The ideal process may be pair creation with equal and opposite momentum (such as an electron-positron pair) since it has the advantage of avoiding any quantum-mechanical ambiguities related to the use of identical particles. In the case of positronium the source radius is about 1 Å, and for a Ca atom about 2 Å. Accumulative interference patterns for one-particle-at-a-time emission has been verified for particles such as photons, electrons, atoms, and neutrons.[6,42,43] In any case, two particles, A and B, are created in the source region simultaneously with approximately equal and opposite momentum. Thus particles B are fired one at a time at the slit plate by repeating the emission process over and over.

As will be demonstrated, triangulation between the spot hit by particle A on screen A, the source and a slit, classically determines whether particle B entered a given slit, or no slit at all. Figure 1 represents both point-like sources and extended sources where the particle-pair is emitted symmetrically with respect to the source center. As shown in Figure 1, if the spot on screen A is in the particle acceptance region to the right of the axis of symmetry, particle B entered the left slit; and vice versa. Observation of the spot hit by particle A permits determination of the trajectory of particle B. If the spot on screen A is not in the particle acceptance region, then it could not have classically gone through either slit. Most classically allowed trajectories will miss both slits. Particle A carries mirror-image information of particle B's trajectory to the slit plate.

**4.1** ***Ratiocination***

In the interference region, the wave function for one-particle experiments, the center-of-mass motion of particle B to the screen B is the sum of two terms due to the two slits

$$\Psi(R,\theta) = \frac{1}{\sqrt{2}}[\psi_1(R,\theta) + \psi_2(R,\theta)], \tag{18}$$

where R is the radial distance to any point on screen B as measured from the center point between the two slits, and $\theta$ is the polar angle. The probability density of a succession of particles B hitting the screen is

$$\Psi\Psi^* = \tfrac{1}{2}\left[|\psi_1|^2 + |\psi_2|^2 + \psi_1^*\psi_2 + \psi_1\psi_2^*\right], \qquad (19)$$

where the interference pattern is due to the cross-terms

$$\left[\psi_1^*\psi_2 + \psi_1\psi_2^*\right] = 2(I_1 I_2)^{1/2}\cos\phi, \qquad (20)$$

where $I_1$ is the intensity of the wave from hole 1 when hole 2 is blocked off, $I_2$ is the intensity of the wave from hole 2 when hole 1 is blocked off, and $\varphi$ is the phase difference between the two waves. The interference pattern, i.e. these cross-terms, are not affected by measurement of the spot position of the twin particle A on screen A if particles A and B do not interfere with each other quantum mechanically or otherwise on their journeys to their respective screens. It may be possible to accomplish this using dissimilar particles A and B.

Because of quantum mechanical entanglement, the two-particle case is more complicated. In the Dirac bra-ket notation, the state of the two-particle system can be written $|\psi\rangle = \frac{1}{\sqrt{2}}\left[|a_\ell\rangle_A |b_r\rangle_B + |a_r\rangle_A |b_\ell\rangle_B\right]$, where the lower case b in the state-vector bracket denotes the path (flight direction) taken by particle B either toward the left slit, subscript $\ell$, or the right slit, subscript r; and similarly for the paths $a_\ell$ and $a_r$ for particle A to screen A.

Unlike the Einstein recoil determination,[7-9] violation of the uncertainty principle can be avoided. Our screens and slit plates may be arbitrarily massive and rigid, so that negligible motion is imparted by momentum transfer of particle A. In addition, screen A may be put so far from the source that particle B interacts with the slit plate and screen B long before particle A impinges upon screen A. Thus neither $\psi_1$ nor $\psi_2$ can be affected as the spot on screen B is determined well in advance of any momentum transfer by particle A on screen A in a massive, rigid system. Future considerations will focus on how much tolerance may be allowed for differences in momentum of the particle pair; and extended sources which emit the particles asymmetrically with respect to the source center. There is a coherence length constraint which limits the size of the source region so that the interference pattern is not lost due to phase differences between the particles as they are sequentially emitted. In an actual experiment, this probably can be met by using laser beam confining and/or laser cooling together with ion traps.[44] Experimentally, for widely separated slits compared to the wavelength of the incident particle, the interference fringes may be difficult to distinguish. In this case sophisticated techniques could be used such as a time-varying shielded magnetic field to periodically vary the interference pattern via the Aharonov-Bohm effect.

The logic of our slit determination is as follows. Due to the uncertainty principle, a spot hit by particle A on screen A projects down to the slit plate as a virtual beam, so that a point on screen A does not translate to a point on the slit plate, but rather as a small area. It will next be shown that the beam spread can be very small compared to the slit width. Figure 1 depicts the case of a very narrow spread of the virtual beam which the analysis justifies. When the spread is a bit bigger yielding a larger $\gamma_R$ region than shown, if particle A hits the region $\gamma_R$, then particle B either misses or enters the right slit, and thus does not at all go through the left slit. If particle B registers on screen B, then it must have gone through the right slit, since it cannot have classically gone through the left slit. A similar consideration applies for spots perpendicular to the plane shown, where the particle appears on screen B and could not have classically gone through a particular slit, so it must have gone through the opposite slit. If particle A hits the region $\delta_R$, then particle B totally misses to the right of the right slit, and if a spot still registers on screen B we have learned a lot that has previously not been known. The inverse holds for the $\gamma_L$ and $\delta_L$ regions. If particle A hits the central region $\alpha$, then particle B has missed both slits hitting the region s between them; but a spot may still register on screen B. By keeping track of the spots and their possible origin, we may determine if all the particles, none, or only subsets produce an interference pattern.

**4.2** ***Compliance with the uncertainty principle***

Let us consider a quasi-stationary source that emits a particle-pair with an uncertainty $\delta x$ in the lateral x-direction parallel to the slit-plate, where $\delta x$ may be considered to be the diameter of the source region in Figure 1. The uncertainty principle gives the uncertainty in lateral momentum of the emitted particles as

$$\delta p_x \geq \frac{h}{4\pi\delta x}, \qquad (21)$$

where h is Planck's constant. The lateral displacement of particle B at the slit plate is

$$\Delta s_x = \left(\frac{\delta p_x}{p_y}\right) D_B, \qquad (22)$$

where $p_y$ is the component of momentum perpendicular to the slit plate, and $D_B$ is the distance from the center of the source to screen B.

To a good approximation

$$\frac{h}{\lambda} = p \approx p_y, \qquad (23)$$

where $\lambda$ is the de Broglie wavelength of particle B.

Thus there is a virtual beam from each spot on screen A to the slit plate. The beam width at the slit plate may even be less than the variable $\Delta s_x$, but it is never greater than $\delta x + 2\Delta s_x$. Figure 2 illustrates the latter case. We want the beam spread at the slit plate from a given spot on screen A, as determined by the projection of $\delta x$ and the momentum uncertainty, in bringing particle B to the slit plate to be small compared with the slit separation

$$s > \delta x + 2\Delta s_x \qquad (24)$$

where s is the distance between the nearest edges of the slits.

Substituting eqs. (21) - (23) in (24), we obtain the quadratic equation

$$(\delta x)^2 - s\delta x + \frac{\lambda D_B}{2\pi} < 0, \qquad (25)$$

whose solution is

$$\delta x < s\left[\frac{1}{2} \pm \frac{1}{2}\left(1 - \frac{2\lambda D_B}{\pi s^2}\right)^{1/2}\right]. \qquad (26)$$

The uncertainty principle is violated if the second term of eq. (26) becomes imaginary. This is easily avoided, and our requirements are readily met if

$$\lambda < \frac{\pi s^2}{2D_B} . \qquad (26)$$

There is no great problem in meeting the condition given by eq. (10) to avoid a violation of the uncertainty principle. However, $\lambda$ should not be made too small, as this would cause the interference fringes to be too close together to distinguish them. In principle, not only can we make the spread of the virtual beam less than the width between slits, we can make it much, much less than their separation.

Superficially it may seem that our experiment is equivalent to the single-slit plate experiment of Einstein analyzed in Section 2.1.2. However, there are some important differences. When the single-slit plate is held rigidly, and an interference pattern results, it is like our source. A momentum transfer measurement is not made on our source to determine particle B's trajectory, as must be done with the plate to obtain trajectory information. This is how violation of the uncertainty principle is avoided. In fact, we need measure neither momentum, nor velocity, but only position and direction. Furthermore, in any of the gedanken experiments of Section 2, the distance from the two-slit plate to the interference screen enters into both the kick and the wave determinations to inextricably lead to a violation of the uncertainty principle. Our experiment avoids this. Finally, our measurements need only be precise on a scale small compared with the slit separation. Because the other experiments rely on the momentum recoil measurement of the single-slit plate, the double-slit plate, or the screen, they unavoidably involve the much smaller scale of the separation of the interference fringes.

The analysis in this section justifies that a very narrow virtual beam (as shown in Fig. 1) is possible which can yield the particle position quite accurately. Future work will consider the ramifications of a broad virtual beam, which may only be realizable in practice. The collision of particle A with screen A will result in highly localized spots. To illustrate correlations between particles A and B, let us consider collisions in the plane of the figure. If these spots form in the designated Particle Acceptance Regions of width $\gamma_R$ and $\gamma_L$, the classical trajectory of particle B enters one of the slits. If particle A hits region $\gamma_R$, particle B should go through the right slit. If particle A hits region $\gamma_L$, particle B should go through the left slit. By triangulation between the spot hit by particle A on screen A and the center of the source, an approximate determination can be made of the trajectory taken by particle B.

For determination of emissions that miss the slits, the slit plate itself could be a luminescent screen. This alone would not yield trajectory information, if done independently of our experiment, in the event that particle B were acted upon by a quantum potential causing its flight path to deviate from a straight line, or by conventional quantum mechanics. However, a luminescent slit plate in conjunction with our two-particle determination could indeed detect the action of a quantum potential or other non-classical effects if the spot on the slit plate were not diametrically opposite the source and the spot on screen A, in deviating significantly from its classically expected position. Making the entire slit plate a detector, greatly adds to the information that can be gleaned from our two-particle, two-slit experiment. If a spot on the slit plate is significantly not in line with a spot on screen A and the source region, then either particle B, or particle A, or both did not follow straight line trajectories.

Motion of particles A and B in opposite directions by conservation of momentum, together with no loss of information carried by particle A about both particles, makes possible classical determination of no slit, or which slit particle B entered. This is the same stratagem used by EPR to circumvent the uncertainty principle to get information about a particle from a measurement made on a partner with which it is perfectly correlated. Determination of which slit can be made either before, at the same instant, or after particle B goes through a slit, depending on the ratio $D_A/D_B$. Non-classically, non-straight line, pathological trajectories in free space are also possible, but they have much smaller probabilities than straight line trajectories. As such they may obfuscate only a small portion of the measurements. We think that our gedanken experiment is not only doable in principle, but can actually be done. Experiment will decide whether or not the interference pattern is destroyed in our "which path" determination.

## 5. Theoretical Attempts to Resolve Wave-Particle Duality

### 5.1 *Bohm's hidden variables*

In Bohm's quantum mechanical theory, there is no wave-particle duality.[45] For Bohm, the particles shot at the slit-plate have definite trajectories, and each particle goes through only one slit or the other. In this theory as excellently presented by Holland[46], the interference pattern results from the interaction of each particle with the quantum potential determined by its own wave function and the presence of the two slits.

If there is a kink in the armor of quantum mechanics, the experiment of Section 4 can shed light on whether the conventional quantum mechanical view, or Bohm's view is closer to physical reality. The aspect of this experiment in which a luminescent slit plate is used to achieve triangulation between a spot on the plate, the source, and screen A, will shed light on whether the trajectories approaching the slit plate are non-classical, i.e. non-straightline.

### 5.2 *Prosser, and Wesley's Poynting vector particle guidance*

In 1976 Prosser made a ground-breaking suggestion that, at least for the case of light, the underlying causal reality for the formation of interference and diffraction patterns is the energy flow given by the Poynting vector.[47] He presented diagrams of the energy flow by a semi-infinite plate with a straight edge, and through two finite-width slits in a slit plate. His solutions appear limited to highly conducting thin plates. Because his illustration was with rather large slits compared to their separation, and the flow is shown only near one of the slits, the figure does not make a convincing case that it can yield the characteristic two-slit interference pattern.

In Prosser's next paper, also of 1976, he describes "wave packets" to represent particles that, while filling all space, could act as localized point particles in certain situations.[48] He suggests a two-slit experiment in which the slits are opened sequentially to test his ideas. He discusses the wave-particle duality and concludes with a new interpretation of the double slit experiment in which photons which pass through the left slit always arrive in the left part of the screen, and no photons go into this area via the right slit; and vice versa. He then compares his interpretation with the Copenhagen interpretation.

In 1984 Wesley[49] independently formulated a similar theoretical concept of the role of the Poynting vector in two-slit interference. Wesley gave due credit to Prosser, and referenced his two papers. He pointed out that smaller slits with wider separation would more clearly show the flow needed to explain two-slit interference. His Fig. 1 clearly shows this, and is remarkably like Fig. 3.1, p. 33 in the recent Bohm and Hiley book.[45] Although they do not mention this similarity, they do reference Prosser and Wesley (p. 269) and do refer to P&W's particle and energy flux suggestions (p. 234).[45]

Wesley associates particle density and flux with the classical wave energy density and flux. Wesley is critical of both quantum theory and of the superposition principle. He notes that "a principle that provides predictions does not necessarily imply causality. Just as one can predict the rising of the sun from the rooster's crow does not mean that the rooster causes the sun to rise." He concludes that "the superposition principle is merely a mathematical convenience devoid of any direct physical or causal significance."

Wesley[49] argues that the quantum potential (such as that of de Broglie, and Bohm) is a strictly "fictitious potential" with no informational content not already present in the particle's motion. For him, "It merely offers an alternate method for representing the particle motion. ... Once a particle is placed in positions with zero velocity, it remains fixed at these positions for all time." He goes on to point out that this is reminiscent of the Bohm theory where all bound particles are motionless. He concludes: "Although the classical wave-particle problem is resolved here by showing how point particle motion can yield 'wave' behavior exactly, it does not really allow one to say that particles are actually involved." For him the underlying reality could be a wave or a flux of particles.

Prosser explicitly states that his solution is for perfectly conducting thin plates. This is only implicit in Wesley's analysis. Their approach would take on much more significance if their results could be generalized to plates made of any material. Even then, there would still be a huge gap explaining interference effects for neutral particles such as neutrons and atoms.

### 5.3 *Davidson's wave-particle duality origin in radiation reaction*

In 1979 Mark Davidson[50] (MD) developed a model in which the radiation reaction force (RRF) is responsible for the stochastic (statistical) origin of Schrodinger's equation. The inspiration for stochastic models stems from the Einstein-Bohr debates of the 1920's and 1930's over the interpretation of quantum mechanics. He points out that in Bohr's interpretation (which is the more widely accepted), "given a physical state, then there is a state vector of some Hilbert space which describes this state completely, but only statistical properties about the physical system can be deduced from this presumed complete description." MD argues that Bohm's hidden variable theory and most others require nonclassical forces (a quantum mechanical potential) to be consistent with Schrodinger's equation which makes them unconvincing.

MD provides a classical explanation of the quantum potential, and the basis for the non-locality of quantum theory by means of the radiation reaction force (RRF). He says, "The quantum mechanical potential implies an unusual force, which acts on the particle, but which depends on the statistical properties of an ensemble of particle trajectories. ... Indeed, it is this extra potential term which leads to quantum interference effects, and the difficulty of describing quantum interference in terms of classical statistical theories has been forcefully stated by Feynman [cf. ref. 51]. ... Preacceleration associated with this radiative reactive force was not considered by Feynman in his arguments."

It is well known that there is a strange aspect to the theory of the RRF in that for a brief period of time it violates causality. Because the RRF is proportional to the time rate of acceleration of the charged particle, i.e. the dynamical equations of motion require a solution of a third order differential equation, it is possible for the charged particle to accelerate for a very short time before a force is applied. The acceleration at any particular time depends on the force to be applied for all future time weighted by a rapidly decaying exponential. There is a typo-graphical error in MD's paper in which he gives this preacceleration time to be $\approx 10^{-22}$ sec for an electron. The correct value is $6.26 \times 10^{-24}$ sec $\approx 10^{-23}$ sec. It is hard to see physically how this extremely short time translates into very long times associated with "delayed choice" and "quantum erasure" thought experiments, except that MD does derive the Schrodinger equation using it and random fluctuations.

For MD the preacceleration helps to explain the nonlocal nature of hidden variable models of quantum mechanics. What is troubling is that the preacceleration itself may be unphysical and a deficit of the present formulation of the radiation reaction force. So using it to justify what appears to be an even more unphysical situation in quantum theory may be a non-sequitur. The fact that the RRF applies only to charged particles is also troubling. MD believes that this is not a difficulty, as many if not all known finite mass neutral particles can be thought to be bound states of charged particles.

**5.4 *Marmet's relativistic waveless and photonless two-slit interference***

Marmet[52] uses an original if not peculiar invocation of relativity theory to obtain interference without either waves or photons. He says, "The wave or photon interpretations are not only useless, they are not compatible with physical reality. Waves are simply the relativistically distorted appearance of relativistic coupling between two atoms exchanging energy." For him there is an ultimate relativistic contraction between the source, the slit-plate, and the screen in the rest frame of a signal going at the velocity of light in vacuum, c. Thus he claims to achieve contact interaction between these three entities. This is difficult to comprehend since it does not occur in the rest frame of the apparatus, and he does not transform back to the rest frame. The remainder is also perplexing, but less difficult. For him, the interference pattern is simply the in-phase and out-of-phase interactions between the oscillators in the source and the oscillators in the screen (the detector) as mitigated by the two slits in close analogy to the classical approach to explaining interference. Thus he has the disturbance simultaneously at both slits, as would be the case for a wave.

Marmet's explanation runs into difficulty if the light goes through a medium which slows its velocity down considerably, $v << c$. Then dimensions would not contract sufficiently for contact interaction. The same difficulty would be encountered in explaining two-slit interference for particles with mass such as electrons, neutrons, and atoms. Hence, Marmet may no longer be able to dispense with the concept of wave, while still keeping the concept of phase.

His relativistic contraction to zero length also buys him a simple explanation of the "collapse of the wave function." In traditional quantum theory, when the quantum wave interacts with the screen, the wave function collapses instantaneously with infinite velocity to the detected spot on the screen; no clear explanation is given of how this physically occurs. Marmet's explanation is "Since in frame c [moving with velocity c] the volume is zero, ... [collapse of the wave function] can be instantaneous, everywhere in that zero volume." Needless to say, there are two problems with thîs explanation. One is that the collapse occurs even when the carrier velocity (be it wave or particle) << c and hence the volume >> 0. The other is that even when the velocity is c, after emerging from a slit the velocity vector is not necessarily parallel to the axis between the slit plate and screen.

There are times even at low velocities, when only relativity can explain otherwise paradoxical phenomena such as the Coulomb and Lorentz force interactions between charged particles in different frames, but it is not obvious that the wave-particle duality in two-slit interference is one of these paradoxes. There is, however, one aspect of Marmet's conjectures that is taken up by others. This is that the wave nature of light is only a reflection of the oscillatory nature of the source of radiation, in a deeper sense than that the source produces the disturbance. It is neither explained by Marmet nor the others how, if this is the case, the oscillation of the radiation persists even after it is decoupled from the source -- even after the source is annihilated.

### 5.4 *Suppes and Acacio de Barros photon trajectories*

In 1993 and 1994, Suppes and Acacio de Barros (S&AB) took a different approach in their attempt to resolve the wave-particle duality.[53,54] They claim that their probabilistic particle approach requires no separate concept of wave to obtain interference. Yet in having their expectation density inherit the periodic wave properties of the oscillating source, this idea is very similar to, if not the same as Marmet's.[52] Their concept is also quite close to the prior ideas of Prosser[47,48] and Wesley[49], except that they additionally assume that their absorber or photodetector also behaves periodically. They differ a little with P&W in having their photons follow linear trajectories except for local interaction with matter. As in the case of Marmet (discussed above), it is not clear how S&AB would explain free-space oscillation of radiation decoupled from its source.

Where they really differ with P&W, is that they invoke interference "locally" at the points of absorption by endowing the photons with two states which can either add or cancel. In essence they have two kinds of photons. For S&AB,[54] "Because only the net excess of positive or negative photons is observable, it is appropriate to call the postive and negative photons *virtual* and thus not necessarily individually observable. Although our concept of virtual photon is not the same as that of QED, we expect common features to be present in our subsequent extension of the present work."

In their 1993 paper,[53] S&AB develop a random walk approach to interference claiming that standard wave concepts can come from purely random particle walks. They conclude that individual photons cannot simultaneously traverse both slits, but only a "distribution and its distribution domain." Their 1994 paper goes much further,[54] and comes to the same conclusion that an individual photon can never go through both slits at the same time.

**6. Conclusion**

According to quantum theory, slit plate momentum transfer detection is incompatible with the formation of an interference pattern. However, this leaves quantum mechanics in a quandary. If a photon goes through both slits at the same time, there is little or no momentum transfer to the slit plate compared with a photon traversing only one slit. Thus even separate detection of the momentum transfers (which destroy the interference pattern and imply one-slit traversal), and the interference pattern (which presumably does not permit detection of the momentum transfers and implies two-slit traversal) are logically contradictory. One might think that Einstein lost the interference pattern debates. However, at the very least, he won a moral triumph here (as in the EPR paradox despite an experimental victory by quantum mechanics) because he succeeded in showing the strangeness and seeming inconsistency of quantum mechanics is.

It is extraordinary from a particle point of view that more photons reach the screen when one slit is closed than when both slits are open. It is amazing that within quantum theory, the proposition that a particle either went through one hole or the other hole is contextually meaningless unless we also specify how this is to be determined. When an interference pattern is built up one particle at a time, we must reject the hypothesis that the pattern is due to an interaction between particles -- at least in real time. Suppes and Acacio de Barros[54] notwithstanding, their virtual photon may have to wait a very long time before interacting with another virtual photon. A purely wave interpretation seems equally unreasonable since individual photons interact only in very localized spots on the screen. Now we also have delayed choice and quantum erasure thought experiments where the present is allowed to influence the past. As mind-boggling as it may seem, one may wonder if hidden deep within the viscera of quantum theory is the sanction to allow time to run both forward and backward. For all these things to be right, the world is not only an uncanny place, it is stranger than we can (or may want to) imagine.

According to Bohr[55]: "What really matters is the unambiguous description of its [nature's] behaviour, which is what we observe. The question as to whether the machine *really* feels, or whether it merely looks as though it did, is absolutely as meaningless as to ask whether light is in 'reality' waves or particles. We must never forget that 'reality' too is a human word just like 'wave' or 'consciousness.' Our task is to learn to use these words correctly -- that is unambiguously and consistently."

What better response than that of Einstein,[56] "There is no doubt that quantum mechanics has seized hold of a beautiful element of truth, and that it will be a test stone for any future theoretical basis, just as electrostatics is deducible from the Maxwell equations of the electromagnetic field or as thermodynamics is deducible from classical mechanics. However, I do not believe that quantum mechanics will be the *starting point* in the search for this basis, just as, vice versa, one could not go from thermodynamics (resp. statistical mechanics) to the foundations of mechanics."

It took over half a century since the beginnings of quantum theory, until it was shown in great generality and detailed analysis by Glauber,[57,58] why it allows interference to be seen in photoelectric type experiments. Until then, and perhaps still, most physicists have incorrectly thought that it is impossible to observe the wave nature (interference effects) of photons in photoelectric currents. As Jaynes has pointed out,[27] unfortunately this lesson has to be relearned periodically because of entrenched biases against it. It is all too easy to accept the underlying assumptions and interpretations of quantum theory -- as unintuitive as they may be -- because quantum mechanics does present a very successful computational edifice.

**Acknowledgments**

I wish to express my gratititude to my esteemed colleagues Arthur Cohn and Mark Davidson for helpful discussions, and to Victor Lohr for his help with the figures.

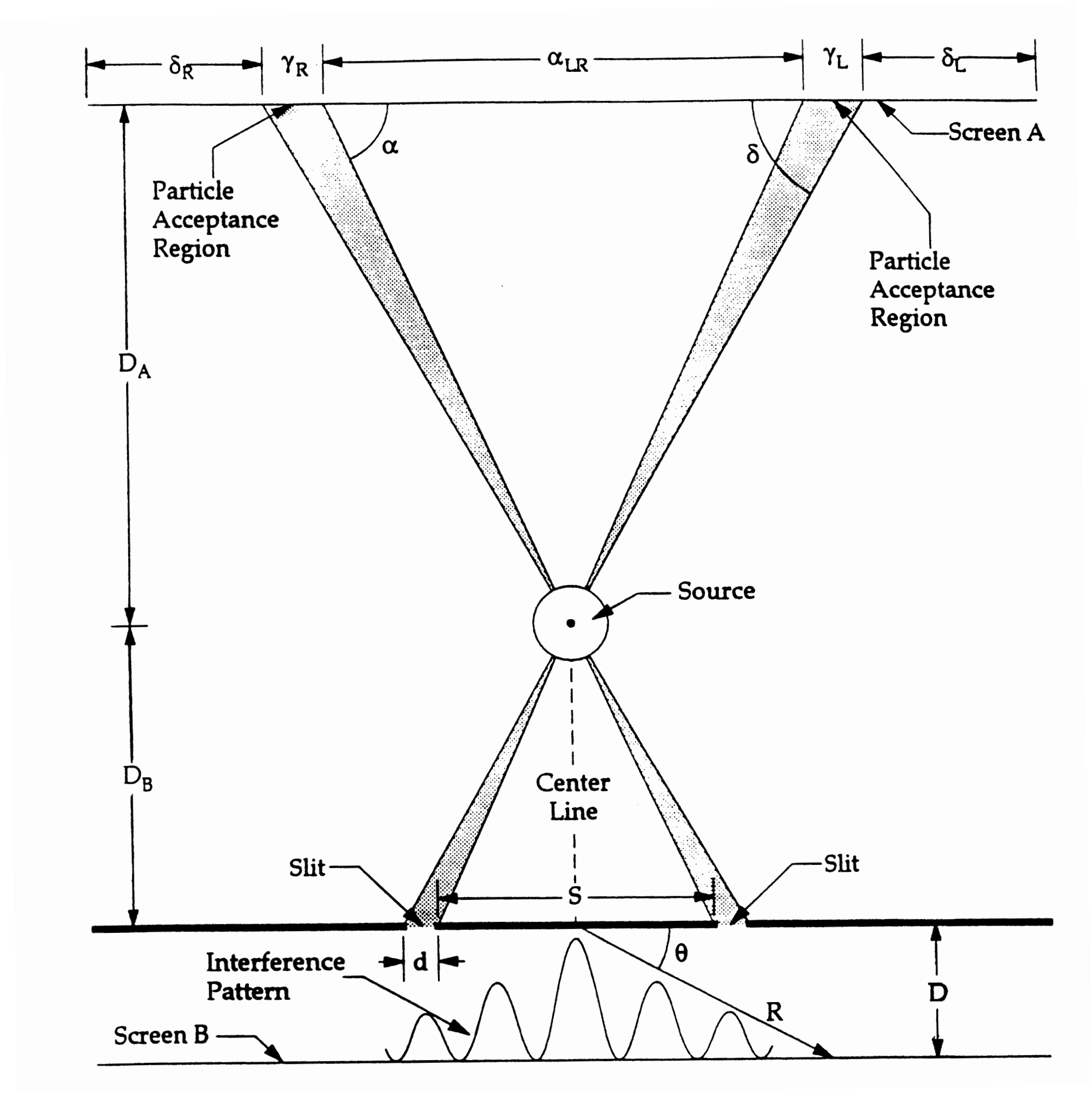

Fig. 1. Two-slit interference experiment in which a particle's trajectory is determined by a partner particle emitted from the source in the opposite direction. Shown are the interference pattern and trajectories (shaded regions) which go through each slit as determined by spots in each of the particle acceptance regions on screen A.

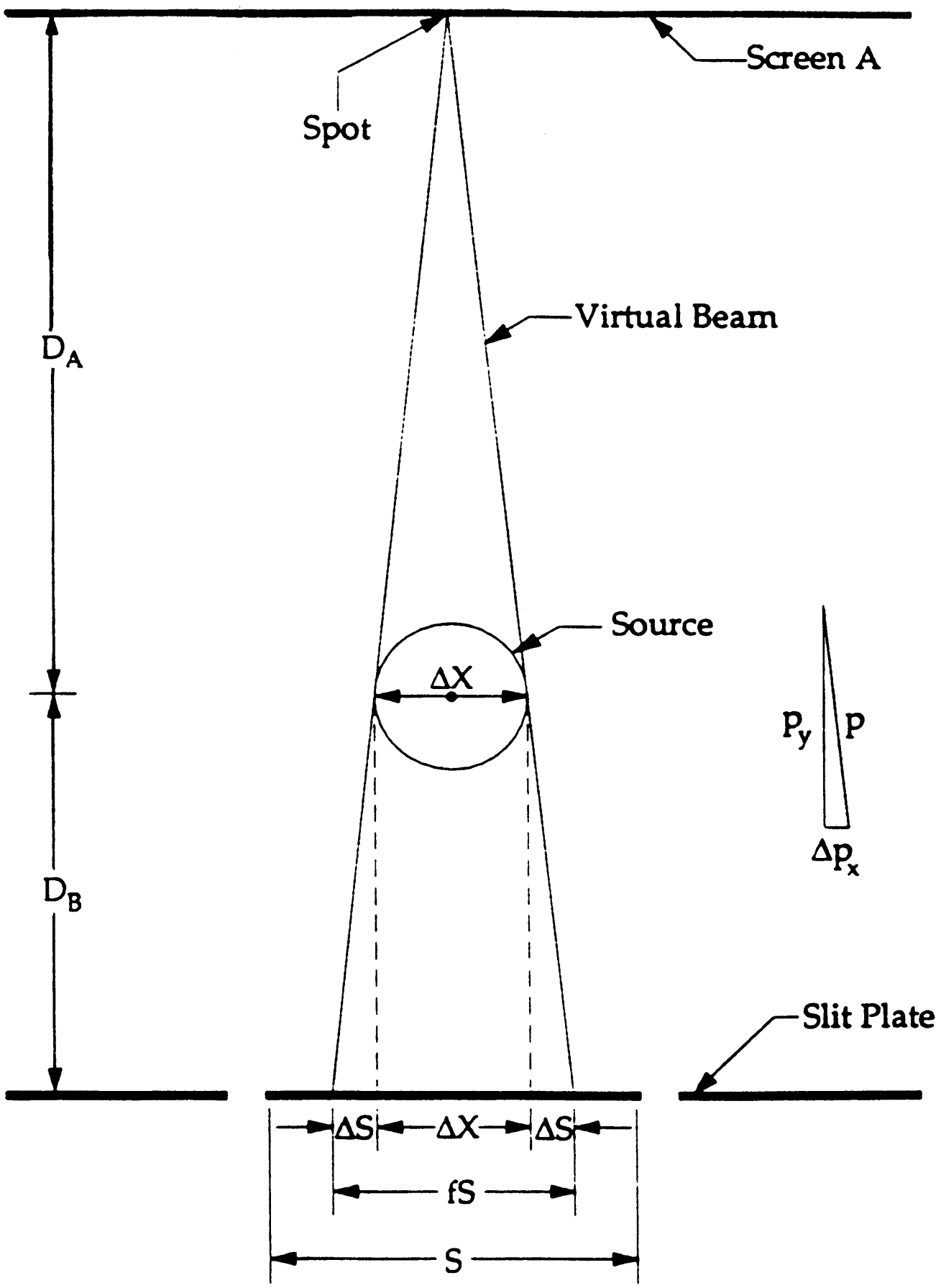

Fig. 2. Diagram used for the analysis that the uncertainty principle is not violated in having a virtual beam of width fS (f is a number < 1), where S is the slit separation. If the spot on screen A is far to the side of the center line, $fS < \Delta x + 2\Delta S$, and may even be less than $\Delta S$.